\documentclass[aps,showpacs,twocolumn]{revtex4}
\usepackage{epsfig}
\begin{document}

\title{The $NN$ phase shifts in the extended quark-delocalization,
color-screening model\footnote{Supported by the National Science
Foundation of China and the Natural Science Foundation of Jiangsu
Province. }}
\author{{LU Xi-Feng$^{1,2}$}, {PING Jia-Lun$^{1,3}$}, {WANG Fan$^{3}$}}
\affiliation{$^1$Department of Physics, Nanjing Normal University, Nanjing, 210097, \\
$^2$National Laboratory for Superconductivity, Institute of
Physics and Center for Condensed Matter Physics, Chinese Academy
of Sciences, Beijing, 100080, \\
$^3$Center for Theoretical Physics and Department of Physics,
 Nanjing University, Nanjing, 210093.}

\begin{abstract}
An alternative method is applied to the study of
nucleon-nucleon($NN$) scattering phase shifts in the framework of
extended quark delocalization, color-screening model(QDCSM), where
the one-pion-exchange(OPE) with short-range cutoff is included.

\end{abstract}


\maketitle

\section{Introduction}
Quantum Chromodynamics (QCD) is generally believed to
be the fundamental theory of the strong interaction. High energy
processes are calculable due to the
asymptotic freedom of QCD. Low energy processes are however
difficult to be calculated directly from QCD, especially for
multiquark systems, due to the infrared confinement and complexity.
At present and even in the future, the
QCD-inspired model will be a useful tool to explore
strong interaction physics at low energy region.

To study the baryon-baryon interaction and multiquark system, the most
popular model is the constituent quark model(CQM). There are
different versions of CQM with different effective degrees of freedom.
The Glashow-Isgur model (GIM)~\cite{GIM,NI}, where the effective degrees of
freedom are constituent quarks and gluons, describes the properties of
hadrons successfully, and has also been extended to the study of the $NN$
interaction. The $NN$ short-range repulsive core is successfully
reproduced, but the intermediate range attraction is missing.
To remedy this shortcoming, two approaches are developed. One
is to invoke meson exchange again, this leads to the 'hybrid'
model~\cite{O,T,F}, another approach is to extend the GIM, this
leads to the quark delocalization color screening
model(QDCSM)~\cite{prl,jlping}. Two approaches both fit the existed
experimental data of the $NN$ interaction, but only the QDCSM explains
the long standing fact that the nuclear force and molecular force are similar
except the obvious differences of the energy and length scales and answers the
fundamental question why does the nucleus is approximately a collection of
nucleons rather than a big ``quark bag". Glozman and Riska argued that the
proper effective degrees of freedom of low energy QCD are constituent quarks
and Goldstone bosons only and proposed the Goldstone boson
exchange model~\cite{GBE}. It also gives a good description of the baryon
spectrum but still in its infancy for the $NN$ interaction~\cite{Stancu}.
This model will have the same drawback as the hybrid model
in the understanding of the $NN$ interaction and nuclear structure.

All of these models enjoy the phenomenological success in the description
of baryon properties and even the baryon-baryon interactions~\cite{hrpang}.
The disagreement with the experimental data can generally be
repaired by fine-tuning of model ingredients and parameters.
However, the more the parameters, the poorer the prediction
power. In this respect QDCSM employs the least adjustable parameters
to explain more physics related to the $NN$ interaction and nuclear
structure. However in the quantitative fit to the  $NN$ and
Hyperon-Nucleon scattering, QDCSM is not as good as the hybrid
model ones up to now~\cite{prl,prc53,npa673}. In order to
explore if the QDCSM can fit the scattering data quantitatively well,
a further extensive calculation is being done.

This paper reports the results of the $NN$ phase shifts calculated
with the extended QDCSM model, where one-pion-exchange
(OPE) with short-range cutoff is added to the original QDCSM. Because the
effective long range $NN$ interaction of the original QDCSM has been found
to be decreased too fast due to a Gaussian wave function used for the
quark orbital motion and so the well established long range $\pi$
exchange is missing. This extended QDCSM reproduces the deuteron
properties very well~\cite{jlping,pang}. In this calculation the
tensor part of one gluon exchange(OGE) is also taken into account.
A new method of calculating the phase shift is tested. Instead of the
usual asymptotic connection to the scattering wave function, we first
solve the equation with a zero boundary condition as one does for the
bound state problem, by varying the boundary to check if the solution
is an exponential decreasing bound state or an oscillating scattering
state, then compare the oscillating wavefunction with that
of free particle to obtain the phase shift. This has been used in the
lattice QCD calculation of the phase shift and so we check
it with this otherwise can be calculated by the normal method.

This paper is organized as follows: in Sect.II. a brief
introduction of QDCSM and calculation method is given, results and
conclusions are given in Sect.III.

\section{Hamiltonian, wave functions and calculation method}
The details about QDCSM can be found in the Refs.\cite{prl,prc51,dstar1}.
Here only the model Hamiltonian, wave functions and the necessary
equations are given. The resonating-group method (RGM), used to
calculate the scattering phase shifts, can also be found in
Refs.\cite{dstar1,Buchmann}.

In QDCSM, the Hamiltonian for the 3-quark system is the same
as the usual potential model and for the 6-quark system, it is
assumed to be
\newpage
\begin{widetext}

\begin{eqnarray}
H_6 & = & \sum_{i=1}^6 (m_i+\frac{p_i^2}{2m_i})-T_{CM} +\sum_{i<j=1}^{6}
    \left( V_{ij}^C + V_{ij}^G +V_{ij}^{\pi} \right) ,   \nonumber \\
V_{ij}^C & = & -a_c \vec{\lambda}_i \cdot \vec{\lambda}_j \left\{
\begin{array}{ll}
 r_{ij}^2 &\qquad \mbox{if }i,j\mbox{ occur in the same baryon orbit}, \\
 \frac{1 - e^{-\mu r_{ij}^2} }{\mu} & \qquad
 \mbox{if }i,j\mbox{ occur in different baryon orbits},
 \end{array} \right. \nonumber \\
V_{ij}^G & = & \alpha_s \frac{\vec{\lambda}_i \cdot \vec{\lambda}_j }{4}
 \left[ \frac{1}{r_{ij}}-\frac{\pi \delta (\vec{r}_{ij})}{m_i m_j}
 \left( 1+\frac{2}{3} \vec{\sigma}_i \cdot \vec{\sigma}_j \right)
  + \frac{1}{4m_im_j} \left( \frac{3(\vec{\sigma}_i \cdot
 \vec{r}_{ij}) (\vec{\sigma}_j \cdot \vec{r}_{ij})}{r_{ij}^5} - \frac{\vec{\sigma}_i \cdot
  \vec{\sigma}_j}{r_{ij}^3} \right) \right], \label{hamiltonian}\\
V_{ij}^{\pi} & = & \theta (r_{ij}-r_0) f_{qq\pi}^2 \vec{\tau}_i \cdot \vec{\tau}_j
  \frac{1}{r_{ij}} e^{-\mu_{\pi} r_{ij}} \nonumber \\
  & & \hspace*{0.2in} \times \left[ \frac{1}{3} \vec{\sigma}_i \cdot
  \vec{\sigma}_j + \left( \frac{3(\vec{\sigma}_i \cdot \vec{r}_{ij})
  (\vec{\sigma}_j \cdot \vec{r}_{ij})}{r_{ij}^2} - \vec{\sigma}_i \cdot   \vec{\sigma}_j \right)
  \left( \frac{1}{(\mu_{\pi}r_{ij})^2} +
  \frac{1}{\mu_{\pi}r_{ij}} + \frac{1}{3} \right) \right], \nonumber
\end{eqnarray}
\begin{eqnarray}
\theta (r_{ij}-r_0) & = & \left\{
\begin{array}{ll}  0 & \qquad r_{ij} < r_0, \\  1 & \qquad \mbox{otherwise},
\end{array} \right. \nonumber
\end{eqnarray}

\end{widetext}
where all the symbols have their usual meaning. OPE potential and the
tensor part of OGE (which is omitted before) have been added.
The OPE was introduced to account for the long range tail of NN
interaction and the tensor part of OGE and OPE were included to do
the S-D channel coupling.
In the OPE potential, $\theta (r)$ is the cutoff function, which is
introduced to avoid double counting in the short range part.
The cutoff parameter $r_0 $ and color
screening parameter $\mu$ are determined by the properties of
deuteron. The quark-pion coupling constant $f_{qq\pi}$ can be
obtained from the nucleon-pion coupling constant $f_{NN\pi}$, which
is determined by experiment.

Combining RGM and generating coordinates formalism, the
ansatz for the 6-quark system wave function can be
written as~\cite{dstar1,Buchmann}
\begin{widetext}
\begin{eqnarray}
\Psi_{6q} & = & {\cal A} \sum_{k} \sum_{i=1}^n \sum_{L_k=0,2} C_{k,i,L_k}
  \int \frac{d\Omega_{S_i}}{\sqrt{4\pi}}
  \prod_{\alpha=1}^{3} \psi_{\alpha} (\vec{S}_i , \epsilon)
  \prod_{\beta=4}^{6} \psi_{\beta} (-\vec{S}_i , \epsilon)  \nonumber \\
  & & \left[ [\eta_{I_{1k}s_{1k}}(B_{1k})\eta_{I_{2k}s_{2k}}(B_{2k})]^{Is_k}
  Y^{L_k}(\hat{\vec{s}}_i) \right] ^J [\chi_c(B_1)\chi_c(B_2)]^{[\sigma]}
    \label{multi} ,
\end{eqnarray}
\end{widetext}
where $k$ is the channel index, $S_i, i=1,\cdots,n$ are generating
coordinates. For example, for the deuteron bound state calculation, we have
$k=1, \ldots,5$, corresponding to the channels $NN~ S=1~L=0$, $\Delta\Delta~
S=1~L=0$, $\Delta\Delta~ S=3~L=2$, $NN~ S=1~L=2$, and
$\Delta\Delta~ S=1~L=2$. $\psi'$s are the delocalized quark orbital wave
functions.
\begin{eqnarray}
\psi_{\alpha}(\vec{S}_i ,\epsilon) & = & \left( \phi_{\alpha}(\vec{S}_i)
+ \epsilon \phi_{\alpha}(-\vec{S}_i)\right) /N(\epsilon), \nonumber \\
\psi_{\beta}(-\vec{S}_i ,\epsilon) & = & \left(\phi_{\beta}(-\vec{S}_i)
+ \epsilon \phi_{\beta}(\vec{S}_i)\right) /N(\epsilon), \nonumber \\
N(\epsilon) & = & \sqrt{1+\epsilon^2+2\epsilon e^{-S_i^2/4b^2}}. \label{1q} \\
\phi_{\alpha}(\vec{S}_i) & = & \left( \frac{1}{\pi b^2} \right)^{3/4}
   e^{-\frac{1}{2b^2} (\vec{r}_{\alpha} - \vec{S}_i/2)^2} \nonumber \\
\phi_{\beta}(-\vec{S}_i) & = & \left( \frac{1}{\pi b^2} \right)^{3/4}
   e^{-\frac{1}{2b^2} (\vec{r}_{\beta} + \vec{S}_i/2)^2}. \nonumber
\end{eqnarray}
$\eta$ and $\chi$ are the spin-isospin and color wave functions.
The delocalization parameter $\epsilon$ is variationally determined by the
six-quark dynamics.

With the above ansatz, the  RGM equation becomes an algebraic equation,
\begin{equation}
\sum_{j,k,L_k} C_{j,k,L_k} H^{k',L'_{k'},k,L_k}_{i,j}
  = E \sum_{j} C_{j,k,L_k} N^{k',L'_{k'}}_{i,j}
   \label{GCM}
\end{equation}
where $N^{k',L'_{k'}}_{i,j}, H^{k,L_k,k',L'_{k'}}_{i,j}$ are the
(eq.(\ref{multi})) wave function overlaps and Hamiltonian matrix
elements, respectively. By solving the generalized eigen value problem,
we obtain the eigen energies of the 6-quark systems and the corresponding
wave functions.

It is well known that the scattering wavefunction of the two-body
scattering process is the spherical Bessel function in the asymptotic
region (if there is no Coulomb interaction or it is neglected)
but has an additional phase shift
in comparing to the free particle ones. So the eigen wavefunction obtained
from the eigen value problem (eq.(\ref{GCM})) of the
scattering state should be proportional to the spherical Bessel function
with a phase shift outside the interaction range.
To fix the phase shift, we compare the nodes of the
obtained wavefunction with the roots of $j_{L}(kR+\delta_L)=0$. In this
way, a set of linear algebraic equations are obtained. The
scattering energy-momentum and phase shift then can be obtained by solving
these equations. In practical calculation, a boundary condition with finite
range (finite R) is always used, so the boundary effect should be
taken into consideration when pick up nodes.

Comparing to the conventional method in obtaining the phase shift, the new
method usually takes more computing time, and doesn't work well for very low
energy scattering because there are not enough nodes in the obtained
wavefunctions. However the new methodhas the advantage that it does not
need to set up an asymptotic boundary, where the matching
conditions are used to find the phase shift. Especially for those cases,
where to really get an asymptotic solution is practically too difficult
and the conventional method cannot be used. The typical case is the
lattice QCD calculation, where one always deals with the problem
in a limited volume. This new method at least is the one to tackle the
scattering for lattice QCD. It is also a convenient method to calculate
the bound and scattering channel coupling problem.

\section{Results and conclusion}
Most of the parameters in the QDCSM  have been fixed by matching
baryon properties, except for color screening parameter $\mu$ and the
short-range cutoff $r_0$. In this calculations, we test three values of $r_0$:
0.6 fm, 0.8fm and 1.0 fm. For each cutoff, $\mu$ is determined by
matching the mass of the deuteron. To save space, here
we only report the results for the best one set ($r_0=0.8$fm)~\cite{pang}
For comparison, the results without OPE are also given. The model parameters
aregiven in Table I.

\begin{center}
Table I. Model parameters
\end{center}

\begin{center}
\begin{tabular}
{|c|cc|cc} \hline & \multicolumn{2}{c|}{without OPE }&
\multicolumn{2}{c|}{with OPE }  \\ \hline $r_0$ (fm) & &
&\multicolumn{2}{c|}{0.8}\\\hline $m$ (MeV) &
\multicolumn{2}{c|}{313} & \multicolumn{2}{c|}{313}\\ \hline
$b$(fm)& \multicolumn{2}{c|}{0.6034} &\multicolumn{2}{c|}{0.6015}\\
\hline $\alpha_s$  & \multicolumn{2}{c|}{1.543}
&\multicolumn{2}{c|}{1.558}\\ \hline $a_c$ (MeV fm$^{-2}$) &
\multicolumn{2}{c|}{25.132} & \multicolumn{2}{c|}{25.135}\\\hline
 $\mu$ (fm$^{-2}$) &
\multicolumn{2}{c|}{1.0} & \multicolumn{2}{c|}{0.85}\\ \hline
$\mu_{\pi}$ (MeV) & & & \multicolumn{2}{c|}{138.04}\\ \hline
\end{tabular}
\end{center}

The following $NN$ phase shifts, $^{3,1}S_{0}$, $^{3,1}D_{2}$, $^{1,3}S_{1}$,
$^{1,3}D_{1}$, $^{1,3}D_{3}$, have been calculated. The symbol
$^{2I+1,2S+1}L_{J}$ is used in the paper and all the experimental data
of $NN$ phase shifts are taken from Ref.~\cite{experiment,www}.
For the case of $^{1,3}S_{1}$, $^{1,3}D_{1}$, coupling between S- and D-wave is
considered.

\begin{figure}
\epsfxsize=8cm \epsfbox{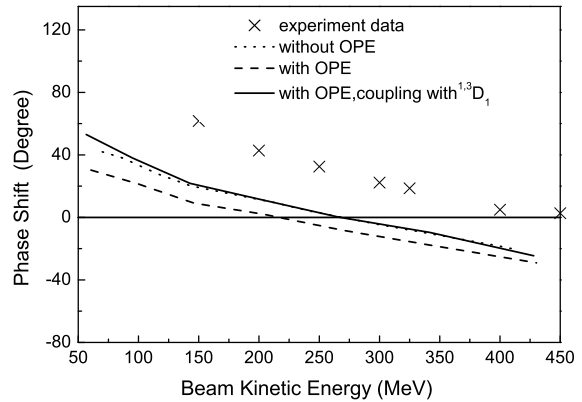} \caption[]
{$N-N$ phase shifts versus laboratory kinetic
energy for the channel of $^{1,3}S_{1}$.}
\end{figure}

\begin{figure}
{\epsfxsize=8cm \epsfbox{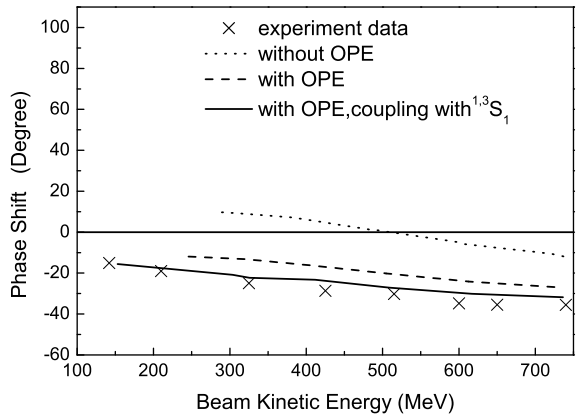}}
\caption[]{Same as FIG.1 for the channel of $^{1,3}D_{1}$. }
\end{figure}

For $^{1,3}S_{1}$ (FIG.1), the phase shifts in the single channel
calculation decrease when OPE is added due to the smaller $\mu$ used.
OPE provided additional attraction and stronger S-D
coupling, the color screening $\mu$ has to be reduced for
compensation to reproduce the deuteron binding energy. Taking into account the
effect of S-D mixing, the phase shifts rise and the results
approach the experiment data again. The remaining discrepancy might be
resolved by additional channel coupling and this has happened in the
bound sate calculation of deuteron.

The $^{3,1}S_{0}$ channel has a similar results
which is not given to save the space.

For the case of $^{1,3}D_{1}$ (FIG.2), OPE  improves the phase
shifts, especially when S-D mixing is considered.
The S-D mixing pushes the phase shifts down, so does the
decreasing of color screening parameter due to the additional attraction
contributed by OPE. The total effect makes the
results match the experimental data very well.

As for $^{1,3}D_{3}$ (FIG.3), the results match the experimental
data as well. As in the $^{1,3}D_{1}$ case, the calculated phase shifts go down
to the experimental data due to the decreasing ofthe color screening parameter
after adding the OPE.

\begin{figure}
{\epsfxsize=8cm \epsfbox{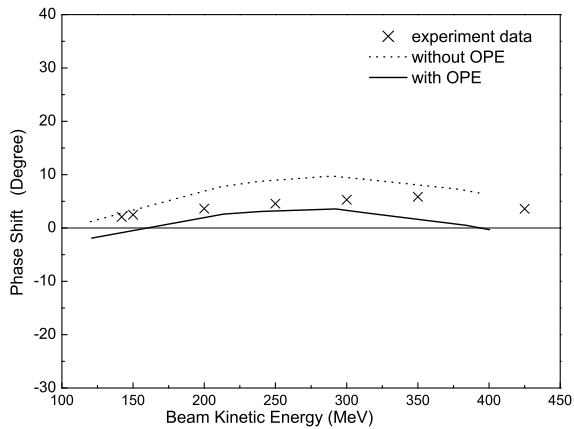}} \caption[]
{Same as FIG.1 for the channel of $^{1,3}D_{3}$.}
\end{figure}

These results show that QDCSM is possible to describe the $NN$ scattering
data quantitatively by including the long range OPE and channel coupling.
Taking into account the following two results together, that the deuteron
properties have been reproduced very well by the same extended
QDCSM~\cite{jlping,pang} and the effective baryon-baryon interactions
of the three constituent
quark models have been shown to be similar\cite{hrpang}, it is fair to
conclude that the short and intermediate range $NN$ interactions,
which were attributed to the $\sigma$ (or two $\pi$)
and heavier meson ($\rho$, $\omega$, etc.) exchanges in the Goldstone boson
exchange and hybrid model, are successfully described by the quark
delocalization and color screening effects.

Quark delocalization is exactly the same as
the electron delocalization which is responsible for the molecular bond.
The nuclear force and molecular force share the same delocalization mechanism
and therefore (at least in our opinion) appear to be similar.
As for the color screening, it is a phenomenology to describe
what has been missing in the two body confinement. Some
nonlinear QCD interactions, such as the three gluon interaction,
three body instanton interaction\cite{LM}, certainly can not be included in
the two body confinement. However a real QCD verification of the model
Hamiltonian assumed in QDCSM is still an expectation.

For the quantitative fit to the scattering data, there are still disagreement
remains to be resolved in QDCSM. The spin-obit interaction of the effective
one gluon exchange should be added to understand the P-wave and higher partial
wave phase shifts. The $N\Lambda$, $N\Sigma$ scattering should be reanalysed
with the Extended QDCSM. The $N\Xi$ scattering should be predicted before
the experimental measurement and it will be a good check of different
constituent quark models.

Our experience on the new method of calculating the phase shifts is
preliminary. Further quantitative check with the usual method within larger
energy region and more channel coupling is needed. Our message to the lattice
QCD phase shift calculation is, it is a unified method to deal with the bound
state and scattering but should be used with caution especially in the
very low energy region and quantitative reliable results being expected.


\end{document}